\documentclass[a4paper,12pt]{article}
\usepackage[utf8]{inputenc}
 \pdfoutput=1

\usepackage{graphicx, float,subcaption}
\usepackage[affil-it]{authblk}
\usepackage[dvips,margin=0.5in]{geometry}

\usepackage{amsmath,stmaryrd} 

\newcommand{\suml}[2]{\sum_{\substack{#1}}^{#2}}
\newcommand{\prodl}[2]{\prod_{\substack{#1}}^{#2}}
\newcommand{\gv}[1]{\ensuremath{\mbox{\boldmath$ #1 $}}} 
\renewcommand{\v}[1]{\mathbf{#1}} 
\newcommand{\avg}[1]{\left\langle #1 \right\rangle} 

\renewcommand{\d}[1]{\mathrm{d}#1\,} 
\newcommand{\rp}[1]{\left( #1 \right)} 
\renewcommand{\sp}[1]{\left[#1\right]} 
\renewcommand{\cup}[1]{\left\{#1\right\}} 
\renewcommand{\cal}[1]{\mathcal{#1}} 
\newcommand{\abs}[1]{\left| #1 \right|} 
\newcommand{\f}[2]{\dfrac{#1}{#2}}
\newcommand{\pd}[2]{\dfrac{\partial #1}{\partial #2}} 
\newcommand{\ra}{\rightarrow}

\newcommand{\g}{\ensuremath{\gamma}}
\newcommand{\s}{\ensuremath{\sigma}}
\newcommand{\w}{\omega}

\newcommand{\vvh}{\v{\hat{h}}}
\newcommand{\vh}{\v{h}}
\newcommand{\vn}{\v{n}}

\date{\today}
\title{Contagion in an interacting economy}
\author{
Pierre Paga\thanks{pierre.paga$@$ kcl.ac.uk}
and
Reimer K\"{u}hn\thanks{Reimer.kuehn$@$ kcl.ac.uk}
}
\affil{
Department of Mathematics,\\King's College London ,UK\\}
\begin{document}
\maketitle

\abstract{
We investigate the credit risk model defined in \cite{HK06} under more general assumptions, in 
particular using a general degree distribution for sparse graphs. Expanding upon earlier
results, we show that the model is exactly solvable in the $N\ra \infty$ limit and demonstrate that the 
exact solution is described by the message-passing approach outlined in  Karrer and Newman \cite{KN10}, generalized to 
include heterogeneous agents and couplings.
We provide comparisons with simulations of graph ensembles with power-law degree distributions.
}

\section{Introduction}

Modern economies form complex, heavily interconnected ecosystems perhaps best epitomized by the extraordinary 
intricacies of supply chains, routinely involving hundreds of suppliers in dozens of countries. 
But as highlighted by Haldane and May \cite{HM11}, complexity is associated with greater systemic
instability, and the crisis of 2007-2009 made it clear that proper analysis of systemic risk is
needed. Accordingly, financial contagion and credit-risk modeling is a long-standing research
subject \cite{Merton74} that has recently attracted renewed interest 
\cite{GK10,ELV07,HK12,Farmer12,MA10}.

Credit events cluster in times of economic stress, resulting in large aggregate losses that are not captured by the risk 
rating (e.g. S\&P) of individual institutions. As a result, attempts at regulatory controls in the spirit of the 
Basel II capital requirements should take into account the possibility of mutually dependent defaults in order to adequately model the risk of large portfolios. 
Historical approaches in financial risk analysis include replacing the number of firms in a portfolio by a reduced 
effective number, or to condition the default probability on macro-economic indicators \cite{Lando,DS99}. 
Multi-factor Merton models correlate defaults by assuming that asset returns of different firms undergo correlated 
random walks \cite{Merton74,elizalde} and that default occurs when the asset level falls below the debt threshold. 
But while these models do take into account correlations, they are not causal and can thus fail to capture systemic 
fragility, i.e. the effect of the collapse of a single entity or group of entities on the entire network.

The physical perspective spurred by the development of econophysics is somewhat different, focusing more on 
system-wide risk through interactions between agents than on individual risks \cite{Durlauf96,Bouchaud13}. 
In particular, there has been in the past twenty years an intensive research effort on the network structure of social and 
economic interactions, e.g. the structure of sexual contacts, of academic citations or of the 
internet \cite{sexweb,citation,www,powergrid}, showing that the number of neighbors or partners of an individual (or 
node) in most social networks follows a power-law distribution. This is sometimes explained as the result of a 
preferential attachment in the network formation process, as in the Barab\'{a}si-Albert model \cite{BA99}.

In this context, a natural framework for systemic risk is provided by contagion processes on graphs defined by economic interactions: 
given such processes, one is interested in the fraction of the network likely to be affected over a certain time horizon.
This approach has been used for example by Gai and Kapadia \cite{GK10,CCD12}, modeling the banking
network 
as a directed weighted network of exposures in which banks fail if they are overexposed to failed
banks. 
The fraction of vulnerable banks can then be determined through a generating function method of the
type used by Moore and Newman \cite{N02,moore2000} to study SIR models on graphs. 
Caccioli et al. \cite{Farmer12} use a representation of the banking system as a bipartite network of
assets and banks where distressed banks sell assets, which lower the price of assets and thus
deteriorates the balance sheet of other banks and precipitates them into distress in turn. 
Hatchett and K\"{u}hn \cite{HK06} investigate contagion in networks of firms via pairwise general
economic interactions that include, but are not restricted to, financial exposures, where default of
neighbors strongly enhance a node's intrinsic probability of default. Contagion on networks, while
a relatively heterogeneous set of phenomena, has nevertheless been largely successfully 
analyzed using tools from statistical mechanics and computer science such as message-passing
\cite{IPC,MP01}.
In the context of infection dynamics for example, Altarelli et al.\cite{HUGEF13-3} evaluate the
efficiency of targeted 
immunization strategies; Karrer and Newman \cite{KN10} derive analytic solution for the
susceptibility of a network to 
SIR-type epidemics.

We focus in this paper on the model developed in Hatchett and K\"{u}hn \cite{HK06}, which
investigates how 
networks of economic interactions affect the system-wide default likelihood across economic cycles.
That 
investigation provided an analytic description of the average fraction of defaulted firms in the
limit of ``dilute yet large'' connectivity, 
i.e. for networks where the average number $c$ of neighbors is large ($c\ra\infty$), but small
compared to the size 
of the system ($\tfrac{c}{N}\ra 0$). Moreover, the analysis was limited to Erd\"{o}s-R\'{e}nyi
random graphs rather 
than more realistic heavy-tailed degree distributions
such as were found by Caccioli et al. for the Austrian banking network \cite{Austrian13}, or by
Souma et al. for links between parent companies and subsidiaries, or between banks and firms in
Japan \cite{SFA03}. 
In what follows we will study networks with heavy tailed degree distributions and finite mean
degrees.

The model has several attractive features: contagion dynamics is not exclusively driven by an
initial shock;
it reveals that systemic risk, while clearly dependent on the system-wide distribution of exposures
and connectivities,
will be relatively insensitive to {\em individual\/} dependencies; the model provides a clear
mechanism for default 
clustering; and (as we shall show in what follows) it is analytically tractable for a larger class
of specifications
than originally considered in \cite{HK06}. Moreover, other models such as the Centola-Macy 
\cite{CM07} or Watts \cite{watts02} models, in which contagion is triggered whenever the number
(resp. the fraction) of contaminated neighbors exceeds a certain threshold, 
can be recovered by taking adequate limits, making it a valuable toy model. Many of its parameters
can also be inferred in principle by suitable rating procedures. Unlike other approaches however 
it does not look into the 'micro-structure' of contagion as generated e.g. by overlapping portfolios
which are the main 
focus of Caccioli et al.\cite{Farmer12}, or by similarities in trading strategies. We believe,
however, that it can be straightforwardly generalized to include such effects at least on a
qualitative 
level and propose a possible approach in the following.

The remainder of the paper is organized as follows: in section 2, we define our model and set up the
formalism
to analyze its dynamics.
In section 3, we provide an analytic solution for general degree distributions (with finite or
infinite connectivity) using recently developed methods in dynamic message-passing, known to be
exact when the underlying network is a tree and generally accurate on a large number of graph
ensembles.
In section 4, we provide numerical results, and summarize our findings in section 5.
In the appendix we use a generating functional analysis to demonstrate that the message-passing
solution is the correct infinite-network limit for a 
configuration model random graph.

\section{Model definitions}
\label{sec:defs}
Our model consists of an ensemble of locally tree-like weighted random graphs $G = (V,E)$ with edge
weights $\rp{w_{ij},w_{ji}}$ drawn according to some distribution $p_w\rp{w_{ij},w_{ji}}$. 
In the context of modeling credit contagion, the nodes describe the entities acting in the economy,
whether they be banks, hedge funds or firms. An edge $(ij)$ can be thought of as a partnership or
joint investment where firm $i$ has a stake $w_{ij}$ and firm $j$ has a stake $w_{ji}$ 
(which needs not be equal to $w_{ij}$). If one of the partners defaults, the other partner writes
its
stake as a financial loss. For example, one may think of a pair of nodes as a supplier of raw goods
and a manufacturer: if the supplier defaults, the need for the manufacturer to find other suppliers
to its specifications has a direct financial effect, while obviously losing a client would be a blow
to the supplier's balance sheet. 

To each node we associate a threshold $\theta_{i}$ and a binary variable $n_{i,t}\in\cup{0,1}$
signifying whether the node is active ($n_{i,t}=0$) or defaulted ($n_{i,t}=1$), the threshold
$\theta_{i}$ corresponding to the assets of node $i$ at the start of the risk horizon. We write
$\v{A} = \cup{a_{ij}}$ for the adjacency matrix of the graph. The degree distribution of the graph
will be kept general in the derivation of analytical results. For numerical results however, a
power-law distribution with exponent $\gamma=3$ and hard cutoffs is used in order to model the
fat-tailed distributions seen in such settings (see e.g. \cite{CCD14}).

In a Merton model, firms are considered as holding debt that must be repaid after a fixed time
period. If their wealth at this time is below the value of the debt, they default, wealth being
modeled as an initial wealth plus a Gaussian random variable. 
As in a Merton model, we assume that firms default when their wealth, i.e. their initial wealth
minus the losses incurred due to the default of their neighbors, plus a random noise term, falls to
zero or below. We write the wealth at time $t$ as
\begin{equation}
\theta_{i,t} = \theta_i + \eta_{i,t} - \suml{j}{} a_{ij}w_{ij}n_{j,t} \ .
\end{equation}
Because of the presence of the noise, defaults are random events. The noise variable $\eta_{i,t}$
has the form 
\begin{equation}
\eta_{i,t} = \xi_{0,t} + \xi_{i,t}\ .
\end{equation}
The $\xi_{0,t}$ are random variables that induce a correlation of the noise across the entire
economy and 
thus represent the effect of varying global economic conditions. Unlike in the usual Merton model
(which uses random walks to determine default probabilities within a one period model), 
we use a multi-period approach in which the $\cup{\xi_{i,t}}$ are idiosyncratic noise variables for
individual time steps. We take the $\cup{\xi_{i,t}}$ to be i.i.d., although we can allow their
distribution to vary with time. 
This decomposition of the noise corresponds to the minimal recommendations of the Basel II Accords
\cite{Basel}. 


The $n_{i,t}$ then evolve according to the dynamical rule
\begin{equation}
\label{eq:1}
 n_{i,t+1} = n_{i,t} + (1-n_{i,t})\Theta\rp{-\theta_{i,t}}\ .
\end{equation}

Written in this way the dynamics have $n_{i} = 1$ as an absorbing state: if $n_{i,t_0}=1$ then for
all $t>t_0$ we have $n_{i,t}=1$. From the modeling point of view, this irreversibility of default is
a reasonable assumption for the limited time horizon considered here. This is a crucial point:
thanks to this absorbing state, we have no memory effects and can obtain $\rho$ easily. Put another
way, we go from having to consider an exponential $(2^T)$ number of possible trajectories for each
node to $T+1$ possible trajectories, resulting in an enormous simplification.

Our aim in this paper is to compute the fraction of defaulted nodes at a finite time horizon $T$,
i.e to compute
\begin{equation}
\rho(T) = \f{1}{N}\suml{i}{}n_{i,T} \ ,
\end{equation}
where $N=\abs{V}$ is the number of firms in the economy. We will call $\rho(T)$ the defaulted
fraction.

Given the way the model is set up, the trajectories $\cup{n_{i,t}}$ undergo Markovian dynamics.
Their joint probability factorizes as
\begin{equation}
P\rp{\cup{n_{i,t}}_{t=0, \cdots,T;\; i\in V}} = 
P(\cup{n_{i,0}})\prodl{t=1}{T}P\rp{\cup{n_{i,t}}_{i\in V}|\cup{n_{i,t-1}}_{i\in V}} \ .
\notag
\end{equation}
In what follows, we will omit giving ranges for the $i$ and $t$ subscripts.

As the $\xi_{i,t}$ are independent, we can integrate over them. The resulting transition
probabilities at time $t$, $P\rp{\cup{n_{i,t}}|\cup{n_{i,t-1}}}$, 
factorize over the sites as
\begin{equation}
P\rp{\cup{n_{i,t}}|\cup{n_{i,t-1}}} = \prodl{i}{}p(n_{i,t}|n_{i,t-1}, \cup{n_{j,t-1}}_{j\in \partial
i}) \ ,
\notag
\end{equation}
in which $\partial i$ denotes the set of neighbors of node $i$. As long as a node $i$ is active at
time $t-1$, 
its transition probability $p(n_{i,t}|n_{i,t-1}=0,\cup{n_{j,t-1}}_{j\in \partial i})$ depends on its
local 
field 
$$
h_{i,t-1} = \suml{j}{}a_{ij}w_{ij}n_{j,t-1} = \suml{j\in\partial i}{}w_{ij}n_{j,t-1} \ ,
$$ 
and on the threshold $\theta_i$, i.e. it will be of the form
\begin{equation}
p(n_{i,t} = 1|n_{i,t-1}=0,\cup{n_{j,t-1}}_{j\in\partial i}) = W_{t-1}\rp{h_{i,t-1}-\theta_i} \ ,
\notag
\end{equation}
where $W_{t-1}$ is a function whose exact form depends on the distribution of the noise
$\eta_{i,t-1}$ at time $t-1$.

Conversely, for nodes that are defaulted at time $t-1$, the single-site transition probabilities are
independent of 
their local fields and are simply given by
\begin{equation}
p(n_{i,t}|n_{i,t-1}=1,\cup{n_{j,t-1}}_{j\in\partial i}) = \delta_{n_{i,t},1}\ ,
\notag
\end{equation}
reflecting the irreversible nature of the dynamics in the sense described above (with $n_{i,t}=1$ as
absorbing state).

Writing $\v{n}_i = (n_{i,0},n_{i,1},\cdots,n_{i,T})$, we find it advantageous to parametrize the
paths $\vn_i$  by 
a single default time $t_i$ defined as the time for which $n_{i,t<t_i}=0$ and $n_{i,t_i}=1$.
Moreover we will assume 
$\cup{n_{i,0}}_{i\in V} = 0$ in the remainder of this article, and omit the dependence on initial
conditions (which 
simply forbid the default time $t_i=0$). Under these assumptions, we can write the path
probabilities as default-time probabilities: 
\begin{equation}
P\rp{\cup{n_{i,t}}} 
= 
P(\cup{n_{i,0}})
P(\cup{t_i}|\cup{n_{i,0}}) 
=\prodl{i}{}P(t_i|\theta_i,\v{h}_i)\ ,
\notag
\end{equation}
where
\begin{align}
\label{def_C}
P(t_i|\theta_i,\v{h}_i) = W_{t_i-1}\rp{h_{i,t_i-1} -
\theta_i}\prodl{s=0}{t_i-2}\sp{1-W_s\rp{h_{i,s}-\theta_i}}\ ,
\end{align}
and where we have introduced the notation $\vh_i = \cup{h_{i,s}}_{s=0,\cdots, T-1}$.

Additionally, we must include the special case where a node does not default within the time horizon
T, corresponding 
to the path $n_{i,t} = 0$ for all $t\leq T$. The probability of such a ``survival'' path of node $i$
is given by
\begin{equation}
\label{eq:survival}
P({\rm Survival}|\theta_i,\v{h}_i) = \prodl{s=0}{T-1}\sp{1-W_s\rp{h_{i,s}-\theta_i}} \ .
\end{equation}
In any sum over default times, the survival path will be implicitly included (it can be
straightforwardly mapped onto 
a default time $t=T+1$ by setting $W_{T}=1$).

The functions $W_t$ encode the noise of the model: in the case where the noise $\xi_{i,t}$ is
Gaussian, which will be our reference, we have
\begin{equation}
\label{Gaussref}
W_t(x) = \Phi\rp{x-\xi_{0,t}}
\end{equation}
where $\Phi$ is the Gaussian cumulative distribution function (cdf), whereas a deterministic model
would have $W_t(x) = \Theta(x)$ where $\Theta$ is the Heaviside function.
\\

We can remark at this point that the $W_t$ need not in general be defined in terms of a cdf and can
be arbitrary transition probabilities, e.g. 
they need not be monotone with respect to the local field. Indeed, whatever the choice of the $W_t$
we remain within the scope of 2-states unidirectional models as defined in \cite{Lokhov14}.

Moreover, bootstrap percolation is recovered by having $P(\v{n_0})$ (or equivalently $W_0$) be a
suitable seeding probability, 
taking constant couplings $w_{ij}= w_0$ and wealth $\theta_i = \theta_0$, taking the zero-noise
limit and setting 
$\theta_0 w_0^{-1} = c$ for a constant $c$ corresponding to the number of defaulted (infected)
neighbors needed to 
propagate default. 
For a Watts-type percolation, $\theta = k\theta_0$ should be considered instead with $k$ being the
node's 
degree. 

In the usual SI model, a node $i$ has a probability per time step $\mu_{ij}$ to be infected by an
infected neighbor, and this can be represented within the current framework by choosing
$w_{ij}=-\log(1-\mu_{ij})$, $\theta = 0$ and $W_t(x) = 1 - \exp\rp{-x}$, along with a suitable
change
in distribution of the $w_{ij}$ to reflect the desired distribution of the $\mu_{ij}$.

\section{Message-passing approach}
\label{sec:MPA}

The following approach is an extension of a method proposed by Karrer and Newman \cite{KN10},
adapted to apply to systems with bond disorder. 
Simpler versions of the resulting equations for homogeneous systems (no bond disorder and uniform
degree) appear in Ohta and Sasa \cite{OS10}. In Altarelli et al. \cite{HUGEF13} and Lokhov et al.
\cite{Lokhov14}, the equations are derived for the single-instance case (without graph ensemble
averaging) in the presence of bond disorder. 

The result that includes graph ensemble averaging, and thus provides a description valid in
the thermodynamic limit, can also be derived from generating functional analysis. In that
derivation, the message-passing equations derived below appear as self-consistency equations in a
saddle-point evaluation of a path-integral. Their detailed structure is a natural consequence of the
dilute nature (relative to system size) of the connection pattern. This alternative derivation is
provided in the appendix.\\ 

We consider a node $i$ of degree $k_i$ and the set of neighboring nodes $\cup{j\in\partial i}$. 
In order to compute node $i$'s default probability at time $t$, we need the marginal default
probability of each of the 
neighboring nodes $j$ at all times $\tau<t$, knowing that node $i$ only defaults at time $t$. 
But given the form of the dynamics, the marginal probabilities of defaulting at a time $t'<t$ do not
depend on the 
specific time of the default $t$, but only on the fact that it is posterior to the default time
$t'$. 
Thanks to this, it is very easy to compute marginals by forward-integration.\\

Consider a specific instance of a weighted tree $G$, and a node $i$ on this graph. We write
$p_i(t_i)$ for the 
probability for $i$ to default at $t_i$. 
If we now consider the neighbors of $i$, we can write that
\begin{equation}
p_i(t_i) = 
\suml{\cup{\tau_j}_{j\in \partial i}}{}
p_i(t_i|\cup{\tau_j}_{j\in \partial i})
\prodl{j\in \partial i}{}
m^{j\ra i}(\tau_j|t_i) \ ,
\notag
\end{equation}
where $m^{j\ra i}(\tau_j|t_i)$ is the probability that $j$ defaults at $\tau_j$ 
conditional on $i$ defaulting at $t_i$. The factorization of the neighbors' conditional
probabilities is due to the 
underlying graph being a tree. In terms of the specifications of the previous section, the
conditional probability 
$p_i(t_i|\cup{\tau_j}_{j\in \partial i})$ is simply $P\rp{t_i|\theta_i, \vh_i}$ defined in
(\ref{def_C}). 

Likewise, we can write
\begin{equation}
\label{eq:conditional}
m^{j\ra i}(\tau_j|t_i) = \suml{\cup{\tau_l}_{l\in \partial j\backslash
i}}{}p_j(\tau_j|t_i,\cup{\tau_l}_{l\in \partial 
j\backslash i})\prodl{l\in \partial j\backslash i}{}m^{l\ra j}(\tau_l|\tau_j)\ .
\end{equation}
But we note that once a node has defaulted, the subsequent dynamics of its neighbors no longer
influence it. In 
particular, the conditional probability $m^{l\ra j}(\tau_l|\tau_j)$ depends on $\tau_j$ only insofar
as 
$\tau_j<\tau_l$. Hence,
\begin{equation}
\forall \tau_j > \tau_l, \qquad m^{l\ra j}(\tau_l|\tau_j) = m^{l\ra j}(\tau_l|\tau_l) \equiv m^{l\ra
j}(\tau_l) \ .
\end{equation}
Likewise, we note that the conditional probabilities $p_i(t_i|\cup{\tau_j}_{j\in \partial i})$ and 
$p_j(\tau_j|t_i,\cup{\tau_l}_{l\in \partial j\backslash i})$ only depend on the neighbors' default
times 
insofar as these precede their own, i.e.
\begin{equation}
p_i(t_i|\cup{\tau_j}_{j\in\partial i}) = p_i(t_i|\cup{\tau_j;\, \tau_j< t_i}_{j\in \partial i})\ ,
\notag
\end{equation}
and it is clear that for all $l,j$,
\begin{equation}
\suml{\tau_l\geq \tau_j}{}m^{l\ra j}(\tau_l|\tau_j) = 1-\suml{\tau_l<\tau_j}{}m^{l\ra
j}(\tau_l|\tau_j) = 
1-\suml{\tau_l<\tau_j}{}m^{l\ra j}(\tau_l)\ .
\notag
\end{equation}
Using these results we can take a new look at the equation for the conditional probability $m^{j\ra
i}(\tau_j|t_i)$ and 
evaluate the r.h.s. of eq.~\eqref{eq:conditional}, expressing it in terms of the $m^{l\ra
j}(\tau_l)$ for 
$\tau_l<\tau_j$. 
For all $r$ in the neighborhood of $j$ ($i$ excepted), we have
\begin{align}
\suml{\cup{\tau_l}_{l\in \partial j\backslash i}}{}
&p_j(\tau_j|t_i,\cup{\tau_l}_{l\in \partial j\backslash i})
\prodl{l\in \partial j\backslash i}{}
m^{l\ra j}(\tau_l|\tau_j)
\notag\\
=&
\suml{\cup{\tau_l}_{l\in \partial j\backslash \cup{i,r}}}{}
\sp{
 \suml{\tau_r<\tau_j}{}
p_j(\tau_j|t_i,\cup{\tau_{l}}_{l\in \partial j\backslash i})
m^{r\ra j}(\tau_r|\tau_j)
\prodl{l\in \partial j\backslash\cup{i,r}}{}m^{l\ra j}(\tau_l|\tau_j)
\right.\notag\\
&\left. \qquad +
\suml{\tau_r\geq \tau_j}{}
p_j\rp{\tau_j|t_i,\cup{\tau_{l}}_{l\in \partial j\backslash i}}
m^{r\ra j}(\tau_r|\tau_j)
\prodl{l\in \partial j\backslash\cup{i,r}}{}
m^{l\ra j}(\tau_l|\tau_j)
}
\notag\\
=&
\suml{\cup{\tau_l}_{l\in \partial j\backslash \cup{i,r}}}{}
\sp{
\suml{\tau_r<\tau_j}{}
p_j\rp{\tau_j|t_i,\cup{\tau_{l}}_{l\in \partial j\backslash i}}
m^{{r\ra j}}(\tau_r)
\prodl{l\in \partial j\backslash\cup{i,r}}{}
m^{l\ra j}(\tau_l|\tau_j)
\right.\notag\\
&\left.\qquad +
p_j\rp{\tau_j|t_i,\cup{\tau_{l}}_{l\in \partial j\backslash \cup{i,r}}}
\rp{
\suml{\tau_r\geq \tau_j}{}
m^{r\ra j}(\tau_r|\tau_j)
}
\prodl{l\in \partial j\backslash\cup{i,r}}{}
m^{l\ra j}(\tau_l|\tau_j)
}
\notag\\
=&
\suml{\cup{\tau_l}_{l\in \partial j\backslash \cup{i,r}}}{}
\sp{
\suml{\tau_r<\tau_j}{}
p_j(\tau_j|t_i,\cup{\tau_{l}}_{l\in \partial j\backslash i})
m^{r\ra j}(\tau_r)
\prodl{l\in \partial j\backslash\cup{i,r}}{}
m^{l\ra j}(\tau_l|\tau_j)
\right.\notag\\
&\left.\qquad +
p_j\rp{
\tau_j|t_i,\cup{\tau_{l}}_{l\in \partial j\backslash \cup{i,r}}
}
\rp{
1-\suml{\tau_r<\tau_j}{}
m^{r\ra j}(\tau_r)
}
\prodl{l\in \partial j\backslash\cup{i,r}}{}
m^{l\ra j}(\tau_l|\tau_j)
}\ .
\notag
\end{align}
Hence we can limit the sum over default times $\tau_l\leq T$ to a sum over default times $\tau_l\leq
\tau_j$, and write
\begin{equation}
m^{j\ra i}(\tau_j|t_i) =
\suml{\cup{\tau_l}_{l\in \partial j\backslash i}\\ \tau_l\leq \tau_j}{}
p_j(\tau_j|t_i,\cup{\tau_l}_{l\in \partial j\backslash i})
\prodl{l\,|\tau_l<\tau_j}{}m^{l\ra j}(\tau_l)
\prodl{l\,|\tau_l= \tau_j}{}
\rp{
1-\suml{\tau<\tau_j}{}m^{l\ra j}(\tau)
}\ ,
\end{equation}
whereas for $\tau_j<t_i$ we have
\begin{equation}
\label{single-instance}
m^{j\ra i}(\tau_j) =
\suml{\cup{\tau_l}_{l\in \partial j\backslash i }\\ \tau_l\leq \tau_j}{}
p_j(\tau_j|\cup{\tau_l}_{l\in \partial j\backslash i})
\prodl{l\,|\tau_l<\tau_j}{}m^{l\ra j}(\tau_l)
\prodl{l\,|\tau_l= \tau_j}{}
\rp{
1-\suml{\tau<\tau_j}{}m^{l\ra j}(\tau)
}\ .
\end{equation}
Replacing $p_j(\tau_j|\cup{\tau_l}_{l\in \partial j\backslash i})$ by its more explicit version 
$P\Big(\tau_j|\,\theta_j,\suml{l\in \partial j\backslash i}{}w_l \v{n}_l\Big)$, this is expressed as
\begin{equation}
m^{j\ra i}(\tau_j) =
\suml{\cup{\tau_l}_{l\in \partial j\backslash i }\\ \tau_l\leq \tau_j}{}
P\Big(\tau_j\big|\,\theta_j,\sum_{l\in \partial j\backslash i}w_l \v{n}_l\Big)
\prodl{l\,|\tau_l<\tau_j}{}m^{l\ra j}(\tau_l)
\prodl{l\,|\tau_l= \tau_j}{}
\rp{
1-\suml{\tau<\tau_j}{}m^{l\ra j}(\tau)
}\ .
\end{equation}
This single-instance equation can be solved by forward propagation from initial conditions for all
nodes, as in \cite{Lokhov14} for a number of models. It has been remarked in several contexts that
message-passing works rather well even if the graph is only locally 
tree-like.
 
Here, we average instead over the degree and wealth of the associated nodes, as well as the 
coupling strengths, to obtain the typical behavior of the system in the infinite system size limit
$N\ra\infty$. 

We note that the neighbors' degree distribution is different from that of a node chosen at random:
the probability 
for a neighbor to have degree $k$ is $\tfrac{kp_k}{\avg{k}}$. Thus the average $m(\tau)$ of $m^{j\ra
i}(\tau)$ satisfies the recursion
\begin{align}
\label{eqp1mp}
m(\tau)  \equiv& \suml{k}{}\frac{kp_k}{\avg{k}}
\suml{\tau_1, \cdots, \tau_{k-1}\leq \tau}{}\;
\prodl{ l|\tau_l<\tau}{}m(\tau_l)
\prodl{l|\tau_l= \tau}{}\rp{1-\suml{\tau'<\tau}{}m(\tau')}
\avg{P\rp{\tau \Big|\,\theta,\suml{l=1}{k-1}w_l\v{n}(\tau_l)}}_{\theta,\cup{w_l}} \ ,
\end{align}
where the average over the couplings $\avg{\cdots}_{\cup{w_l}}$ is done over the marginal coupling
distribution.

The resulting equation is forward-propagating in $m$, starting from $m(1) =
\avg{W_0(-\theta)}_\theta$.
The average fraction of defaults happening at time $t$, meanwhile, is given by
\begin{align}
\label{marginal}
p(t)\equiv&\f{1}{N}\suml{i}{}p_i(t)
\notag\\=&
\suml{k}{}p_k
\suml{\tau_1, \cdots, \tau_{k}\leq t}{}\;
\prodl{l|\tau_l<t}{}m(\tau_l)
\prodl{l|\tau_l=t}{}\rp{1-\suml{\tau'<t}{}m(\tau')} 
\avg{P\rp{t\Big|\,\theta,\suml{l=1}{k}w_l\v{n}_l}}_{\theta, \cup{w_l}} \ .
\end{align}
As mentioned in \cite{KN10}, the equation we obtain is that of a single representative message.\\

The resulting numerical scheme is transparent and can be used to quickly compute average default
probabilities on an ensemble of locally tree-like graphs.
Assuming the message $m(s)$ for $s$ up to $t-1$ have already been computed, the procedure is: 
\begin{itemize}
\item draw a degree $k$ according to the neighbor degree distribution $\tfrac{kp_k}{\avg{k}}$, and a
wealth $\theta$,
\item draw  $k-1$ interaction weights $\cup{w_l}_{l=1, \cdots, k-1}$,
\item draw $k-1$ default times according to the previously computed distribution $(m(0),
m(1), \cdots,m(t-1), 1-\sum_{s<t}m(s))$,
\item compute the resulting $P\rp{t|\theta,\vh}$,
\item repeat the procedure $N_{sampling}$ times and average the results to give $m(t)$ according to
eq.~\eqref{eqp1mp}.
\end{itemize}
The average default probability $p(t)$ can be computed in parallel to $m(t)$ according to
eq.~\eqref{marginal}, by drawing a degree $k$ according to $p_k$, and 
drawing $k$ interactions weights and default times (according to $m$).
The defaulted fraction $\rho(t)$ is then given by
\begin{equation}
\rho(t) = \suml{s=0}{t}p(s) \ .
\end{equation}

This algorithm is thus a form of Monte Carlo sampling, since instead of doing an exhaustive
summation over all possible trajectories (which would have complexity  $T^{k_{max}}$), 
we sample over the default times according to the distribution $\cup{m(s)}_{s<t}$, bringing the
complexity down to $N_{sampling}\times T \times \avg{k}$ where $N_{sampling}$ depends on the
required precision. 

Unlike standard message-passing equations, which take the form of self-consistency conditions and
are thus usually iterated until convergence, the equations are forward-propagated.


\section{Numerical results}
In the following we present results of the above analysis for a stylized economy exhibiting mutual
financial exposures which constitute a graph of dependencies with a power-law degree distribution. 
The analysis presented here can be done with an arbitrary degree distribution with finite mean
degree $\avg{k}$. 
We take a truncated power-law as a relatively straightforward choice allowing us to investigate the
effect of heterogeneity in the degree distribution on default statistics. 
This regime had so far remained unexplored in \cite{HK06}, which was restricted to systems based on
large $\avg{k}$ Erd\"{o}s-R\'{e}nyi random graphs, and thus to unrealistically homogeneous
networks. 
While real markets do not exhibit clear power laws in their degree distributions as noted by
\cite{Austrian13}, heavy tails are a common features and this tail behavior is often reasonably
well modeled by power-law degree distributions (\cite{SFA03}).

In principle, three levels of analysis are available: 
\begin{itemize}
\item[(i)] using population dynamics to study equation~\eqref{marginal} 
\item[(ii)] simulating (large) single instances to solve equation~\eqref{single-instance} 
\item[(iii)] stochastic simulations for a moderately large system sizes to check the validity of the
theoretical 
analysis
\end{itemize}
Single-instances cavity equations have been studied in the case of bootstrap percolation in
\cite{HUGEF13}, and in general contexts in \cite{Lokhov13, Lokhov14} and will thus not be
investigated here. Instead, we will focus on comparing the results of 
$(i)$ and $(iii)$.

Unless otherwise specified, we will use a truncated power-law degree distribution, $p_k\sim k^{-\g}$
with $\g=3$,
for $k\in \llbracket k_{min},k_{max} \rrbracket$ with various values of $k_{min}\ge 1$, and
$k_{max}=100$. The 
wealth $\theta$ will be a Gaussian r.v. with mean $\theta_0 = 2.75$ and variance $\s_{\theta} =0.3$
as used in 
\cite{HK06}. We take $W_t(x) = \Phi(x)$ for all $t$, assuming neutral macro-economic condition
(i.e. 
$\xi_{0,t}=0$ in \eqref{Gaussref}) everywhere except in sections 4.2 and 4.4. The couplings are
taken to be Gaussian r.v. with mean $w_0=1$ and variance 
$\s_w=0.5$ except in section 4.3. For simulations, we take for the network size $N=10^3$. The time
horizon is taken to be $T=12$.

The parameter values used here are primarily for the convenience of obtaining a robust signal
without resorting to networks so large as to make simulations prohibitively time-expensive, while
ensuring the right order of magnitude for typical annual default rates.

\subsection{Initial acceleration}

As shown in \cite{HK06}, we can qualitatively assess the interaction-induced increase in risk by 
looking at the discrete second derivative of the defaulted fraction at $t=1$, $\Delta_1 =
\rho(2)+\rho(0) - 2\rho(1) = p(2)-p(1)$. A 
positive initial acceleration of the fraction of defaulted firms can be seen as an indicator of
destabilization of the 
economy through mutual exposures. Indeed, for a non-interacting system the initial acceleration is
quickly found to be 
$$
\Delta_1 = -\avg{W_0(-\theta)W_1(-\theta)}_\theta<0 \ .
$$

In order to compare the results across networks with different mean degrees and with the high
mean-degree results of 
previous works we plot the values of $\Delta_1$ in the space of interaction parameters $\rp{w_0, 
\s_w}$. However, a higher degree means more liabilities and thus a possibly much higher likelihood
of losses. 
Hence in order to make the results comparable between different degrees the coupling strength
parameters are 
rescaled: as in \cite{HK06} we take  $w_{ij} = w_0\avg{k}^{-1} + x_{ij}\s_w\avg{k}^{-1/2}$, with 
$x_{ij}\sim \cal{N}\rp{0,1}$. 
While the values of the $\Delta_1=0$ boundary depend on the details of the degree distribution, we
find the 
theoretical predictions for large mean degree to agree remarkably well with previous results for 
high-connectivity Erd\"{o}s-R\'{e}nyi random graphs, even in the case of a power-law distribution
with finite average connectivity. 

This is due to a combination of the graph being locally tree-like, i.e. early defaults in the
neighborhood of a node 
are independent, and the initial default probabilities being low (of the order of $10^{-3}$). Taken
together, this allows us to assume that typically at most one among a node's $k$ neighbors defaults
in the first time steps. 
The resulting contribution of the interactions to the acceleration for a node of degree $k$ can be
shown via eqs.~\eqref{eqp1mp} and \eqref{marginal} to be 
$\delta p(2) = k\,m(1)\,\avg{w}_w\partial_{h_1}\avg{P(2|\theta,\v{0})}_\theta$. Averaging over the
interaction distribution yields $\avg{w}_w = \tfrac{w_0}{\avg{k}}$, and once averaged over the
degree distribution the resulting contribution has no dependence on the mean degree.
Hence it is shared among all tree-like graph ensembles, the Erd\"{o}s-R\'{e}nyi ensemble among
them,
and admits a finite limit in the large mean connectivity limit. 
We thus recover the limiting case explored in \cite{HK06}.

\begin{figure}
\centering
\includegraphics[width=0.9\textwidth, clip=true,trim = 0mm 70mm 0mm 70mm]{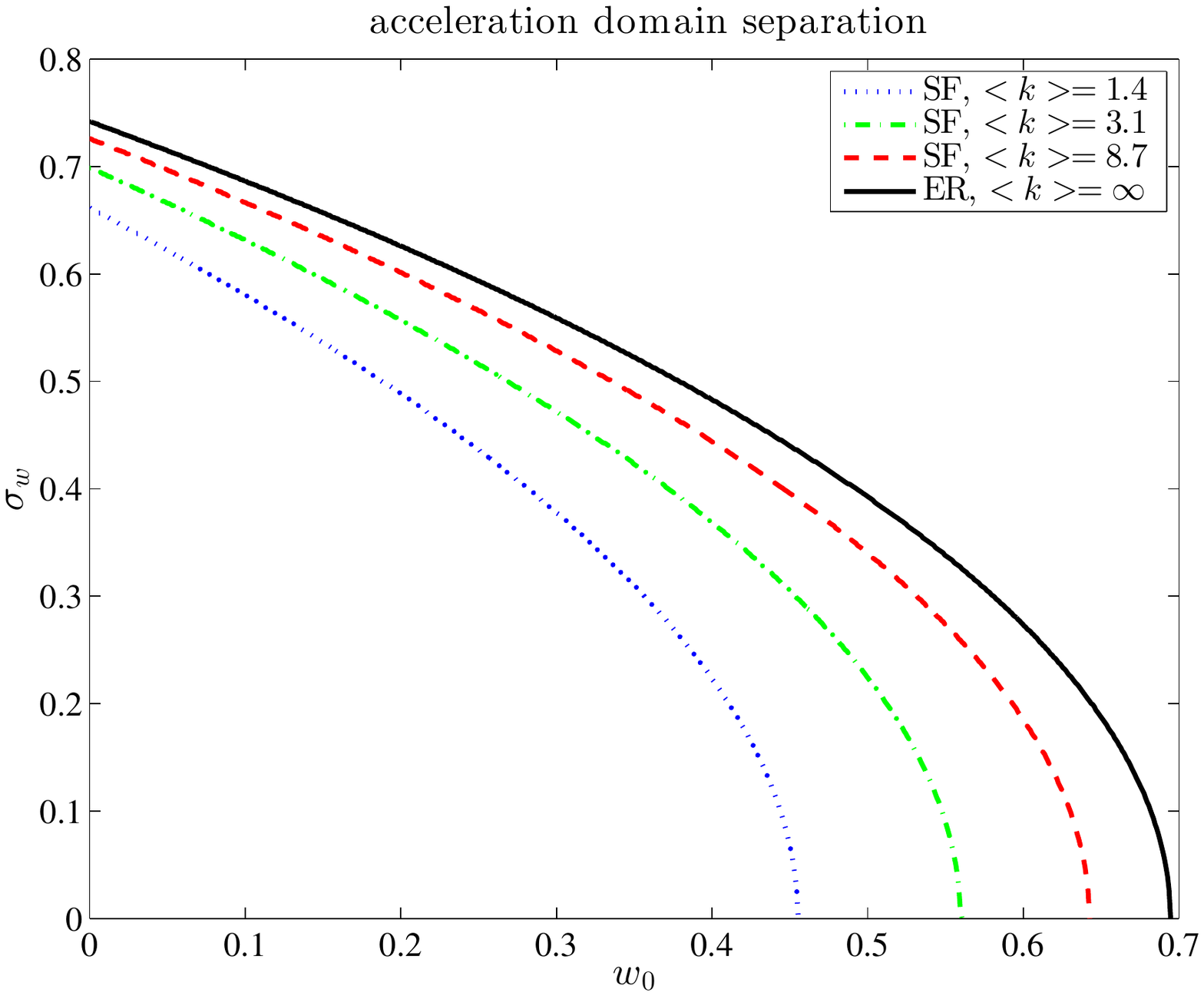}
\caption{
Acceleration domain boundaries for power-law graphs with $\avg{k}\simeq 1.4$ (blue, dotted),
$\avg{k}\simeq 3.1$ (green, dash-dotted) and $\avg{k}\simeq 8.7$ (red,dashed), and the 
Erd\"{o}s-R\'{e}nyi graph with infinite connectivity (black, solid). The external domain (large
$w_0$, large $\s_w$) has positive acceleration, 
marking a destabilizing effect of interactions.
}
\label{fig-acc}
\end{figure}

The domain boundaries are plotted in figure \ref{fig-acc} in the case of a truncated power-law
degree distribution for $k_{min}=1$ (i.e. $\avg{k}\simeq1.4$),
 $k_{min} = 2$ ($\avg{k} \simeq 3.1$) and $k_{min} = 5$ ($\avg{k} \simeq 8.7$).
The Erd\"{o}s-R\'{e}nyi case with infinite connectivity is added for reference.

\subsection{Macro-economic sensitivity}

The Basel regulatory framework (Basel II and III) requires banks to take into account cyclical
effects and 
macro-economic factors in their risk estimate, in the shape of a countercyclical capital buffer. It
is thus worthwhile 
to investigate the default probability across the economic cycle in our setting, highlighting once
again the 
destabilizing effect of interactions. Assuming for simplicity that these cyclical effects to change
slowly over the 
course of a year, we set $\xi_{0,t} = \xi_0$ in \eqref{Gaussref} to reflect macro-economic
condition, and set $\xi_0$ 
to be a Gaussian r.v. (positive $\xi_0$ reflecting favorable, negative $\xi_0$ reflecting
unfavorable conditions).

\begin{figure}
\centering
\includegraphics[width=.8\textwidth, clip=true,trim = 0mm 60mm 0mm 80mm]{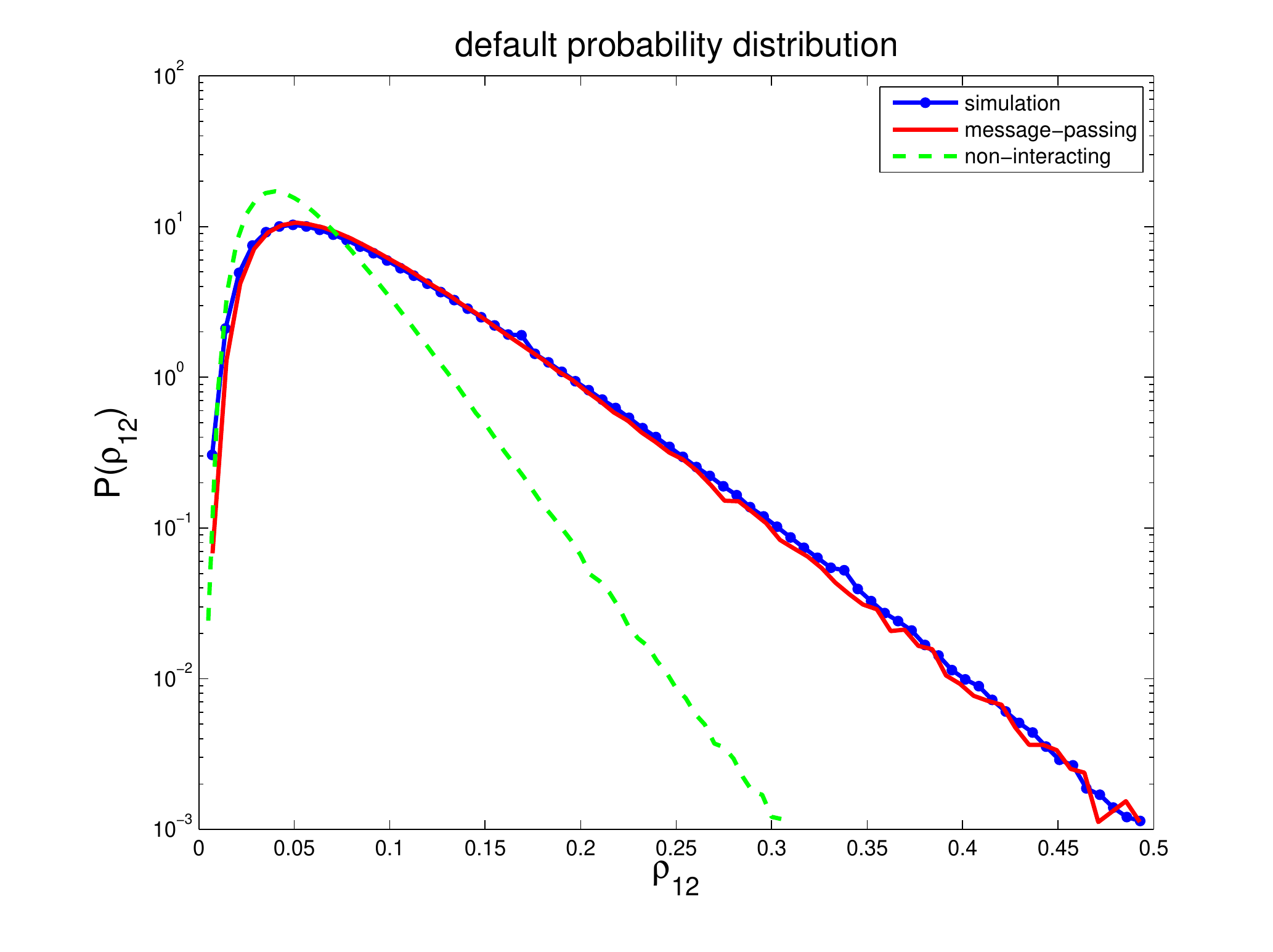}
\caption{
Default distribution for $\xi_0\sim \cal{N}\rp{0,0.2}$: Simulations (blue dotted line) compared with
cavity predictions (red solid line) and 
with the non-interacting network (green dashed line).
}
\label{xi}
\end{figure}

We can then study the distribution $P(\rho_T)$ of the defaulted end-of-year fraction $\rho(T)$
induced by the distribution 
of $\xi_0$, the shape of the large $\rho(T)$-tail giving an indication as to the vulnerability of
the system to large-scale 
economic shocks. This is done in figure~\ref{xi}, with an empirical distribution obtained from
$10^7$ runs ($10^3$ runs on $10^4$ graphs). The non-interacting case is added for comparison. 

As we can see the tail of the probability at large defaulted fraction $\rho$ is noticeably fatter in
the presence of interactions, making the systemic risk much larger in bad macro-economic conditions,
and suggests that correspondingly large buffers are needed.

Within the present study we have not specifically looked into the effects of pro-cyclical or
countercyclical behavior of banks, although such effects could be included in our model. We refer
to the concluding section for a discussion of these effects.


\subsection{Interaction strength}

\begin{figure}
\centering
\includegraphics[width=0.9\textwidth, clip=true,trim = 0mm 60mm 0mm 80mm]{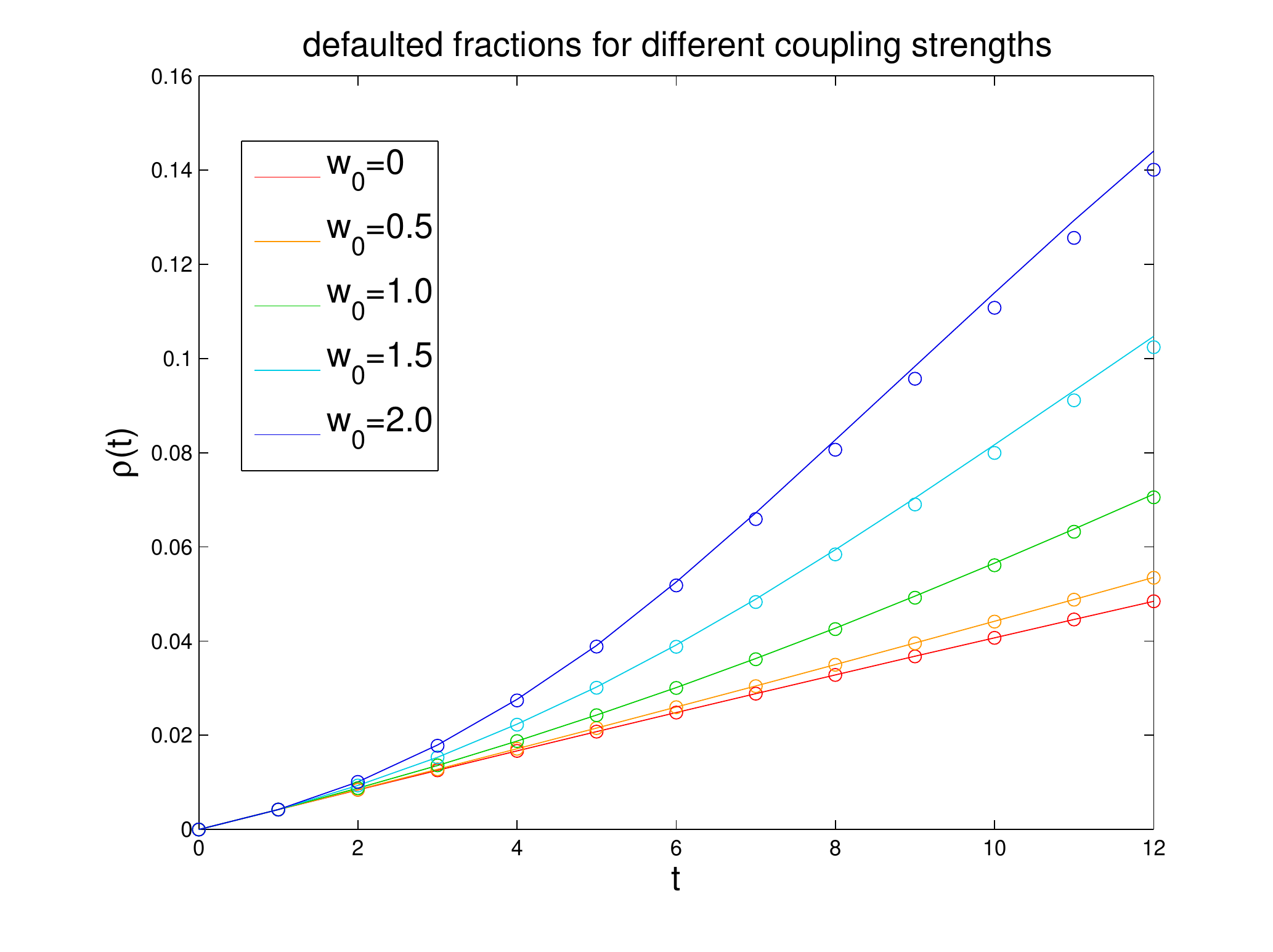}
\caption{
Defaulted fractions for different mean coupling strengths: simulation (circles) and cavity
predictions (solid lines).
}
\label{int-strength}
\end{figure}

In figure \ref{int-strength} we plot the time dependent defaulted fraction $\rho(t)$ at neutral
macro-economic conditions
($\xi_{0,t}=0$) for different values of the interaction strength $w_0$. We set $\s_w$ to half 
the value of $w_0$, and average the simulation results over 250 graphs, with 500 runs on each
graph. 

As expected, the finite-size effects becomes larger as the interaction strength increases, which is
to be expected: 
while the cavity method is exact on trees, the presence of loops strongly affects its performance.
And while the graphs 
used are locally tree-like, for finite size systems there are still a large number of loops
remaining. Since these loops 
only affect the dynamics when their constitutive nodes are defaulted, their effect is felt more
strongly when the default 
rate is higher.

\subsection{Extension: spillover}

An important extension of the model is the inclusion of spillover effects, as induced by asset
fire-sales. 
A fire-sale happens when a firm, short on liquidity, sells a large amount of assets in a short time
(in the 
case of banks, it is a regulatory requirement to maintain a certain size of its liquid capital
buffer). As 
a result of the sudden glut, asset prices will fall, diminishing the value of assets held by other
firms. 

To implement this, we consider for simplicity a single class $a$ of (tradable) assets, representing
a given 
(constant) fraction of every firm's wealth. When a firm's wealth falls below a certain fraction
$f_c$ of its 
initial wealth it enters a distressed state and sells asset $a$ to maintain liquidity, resulting in
a fall of 
the asset price. As a result, the value of the portfolio of \emph{every} firm holding this asset
falls, which we model 
by changing the firms' initial wealth by a factor $r(d_t) =  (1+r_0\,d_t)^{-1}$ where $d_t$ is the
fraction 
of distressed firms at time $t$ (defaulted firms are considered distressed as well), and $r_0$ a
parameter that
could describe market depth. Thus, a firm's wealth at time $t$ becomes
\begin{equation}
\theta_{i,t} = r(d_{t-1})\,\theta_{i} - \suml{i}{}w_{ij}c_{ij}n_{j,t} - \eta_{i,t}\ ,
\notag
\end{equation}
and default is triggered when $\theta_{i,t}$ falls below zero, while the firm enters a distressed
state as soon
as $\theta_{i,t} < f_c \theta_{i}$. 

\begin{figure}
\setcounter{subfigure}{0} 
\centering
\begin{subfigure}{.8\textwidth}
	\includegraphics[width=.8\textwidth, clip=true,trim = 0mm 70mm 0mm 80mm]{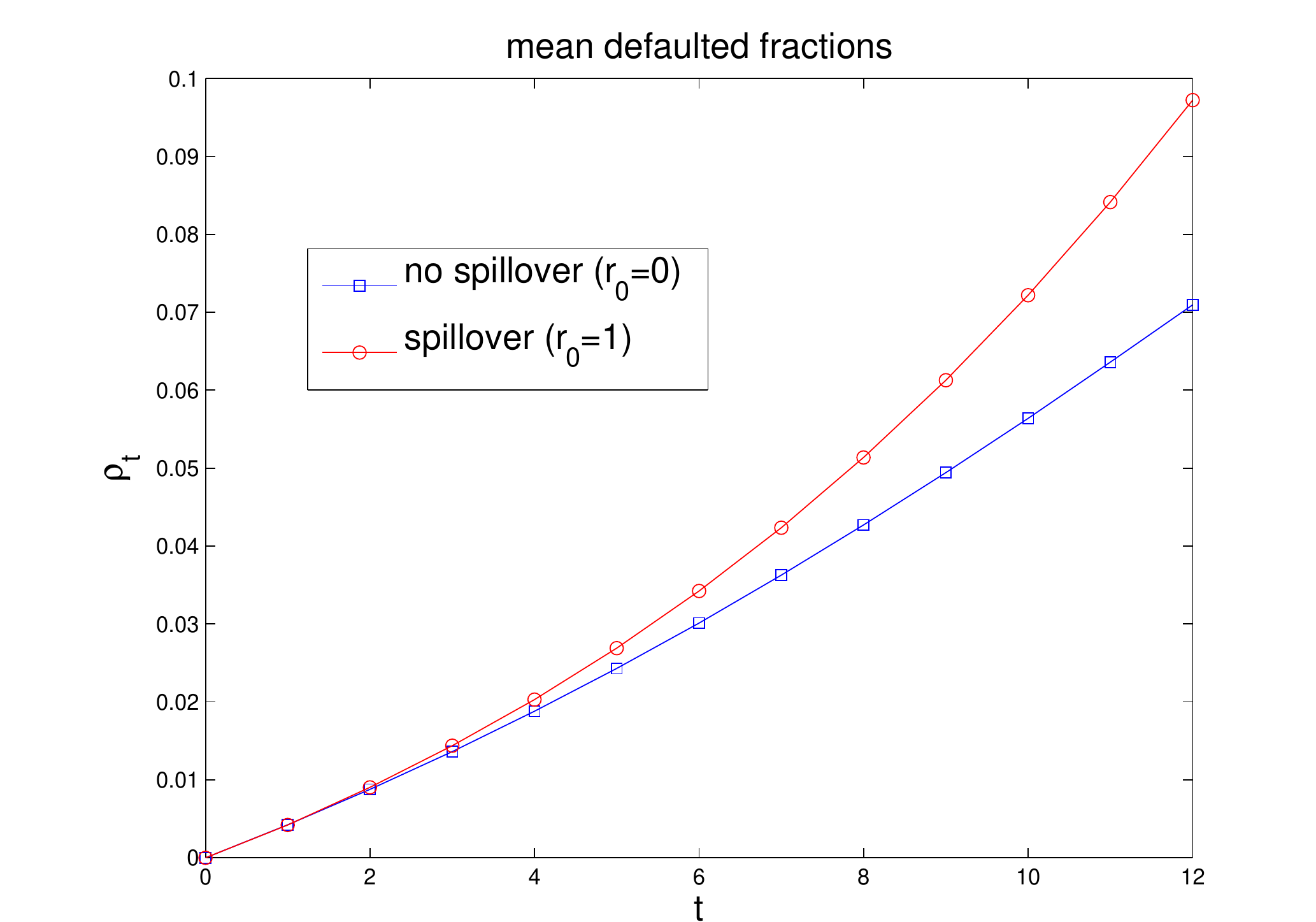}
\caption{
Mean defaulted fractions under neutral macro-economic conditions ($\xi_0=0$) for a model with
spillover  $r_0=1$ 
(red circles) compared to results without spillover (blue squares).} 
these results
\label{fig-spill-a}
\end{subfigure}
\begin{subfigure}{.8\textwidth} 
	\includegraphics[width=.8\textwidth, clip=true,trim = 0mm 70mm 0mm 80mm]{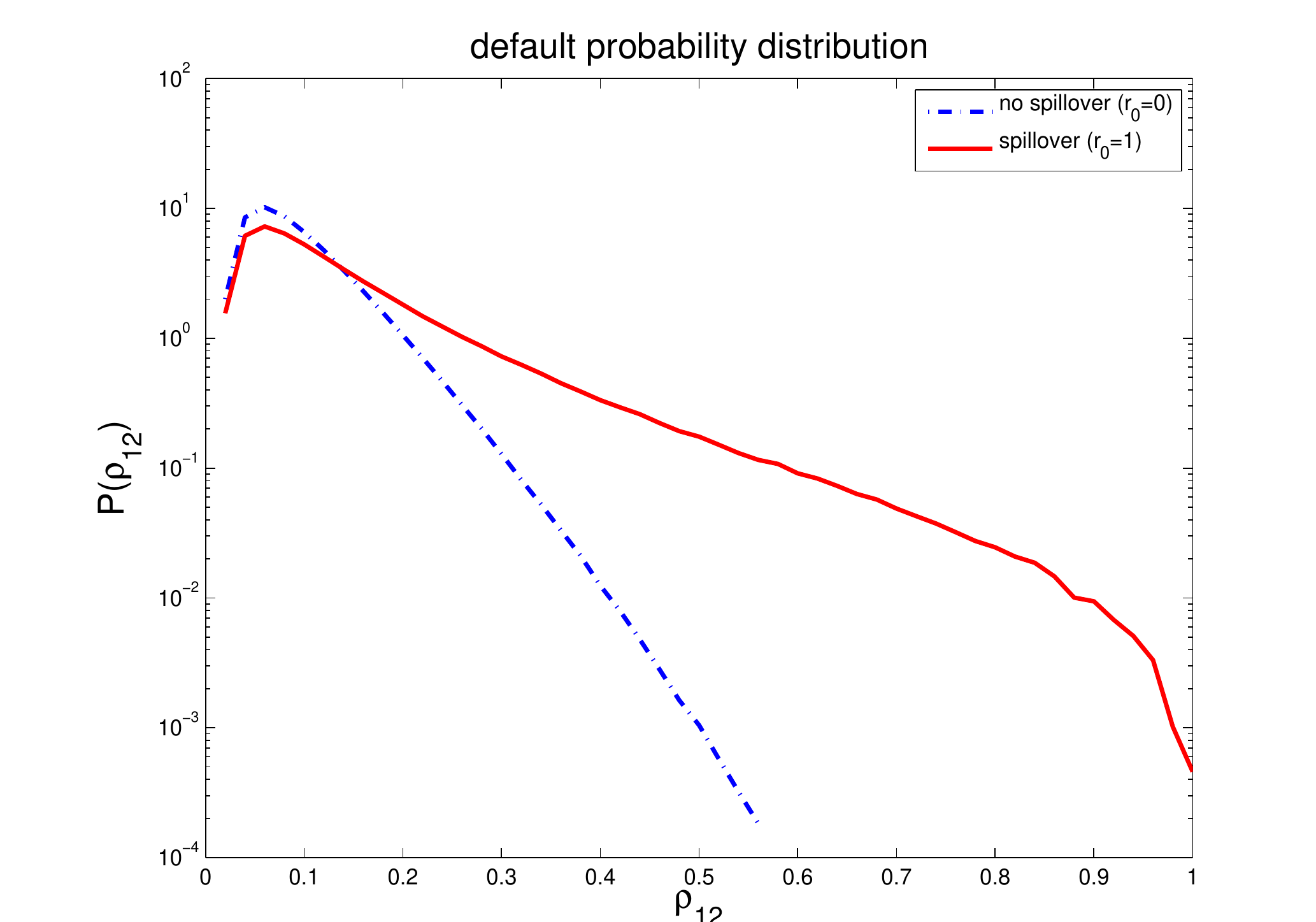}
\caption{
End-of-year default probabilities with spillover $r_0=1$(red solid line) compared to results without
spillover (blue dotted line) for 
$\xi_0\sim \cal{N}\rp{0,0.2}$.
}
\label{fig-spill-b}
\end{subfigure}
\label{fig-spill}
\caption{}
\end{figure}

This can be seen as implementing a simple version of the overlapping portfolio approach of
\cite{Farmer12} on top of the credit risk model, using in the present case only one asset class.
Plotted below are the mean defaulted 
fraction in neutral macro-economic conditions ($\xi_0=0$, fig.~\ref{fig-spill-a}) and the
probability distribution of 
defaulted fractions across the economic cycle as generated by a distribution of macro-economic
conditions (fig.~\ref{fig-spill-b}), showing that spillover effects can 
dramatically increase the probability of large defaulted fractions. We take $f_c = 0.1$.

While the network of exposures induced by assets is very different from the simple pairwise
interaction network due to the inclusion of a bipartite (firms-assets) structure, as long as the
number of asset classes remains finite in the large-system limit their contribution can be modeled
by a simple mean-field effect as above. 
If the number of asset scales with $N$ however, the situation becomes much more complicated and
depends on the details of the network structure and the interaction type. 

\subsection{Network size}

\begin{figure}
\centering
	\includegraphics[width=0.9\textwidth, clip=true,trim = 0mm 70mm 0mm 80mm]{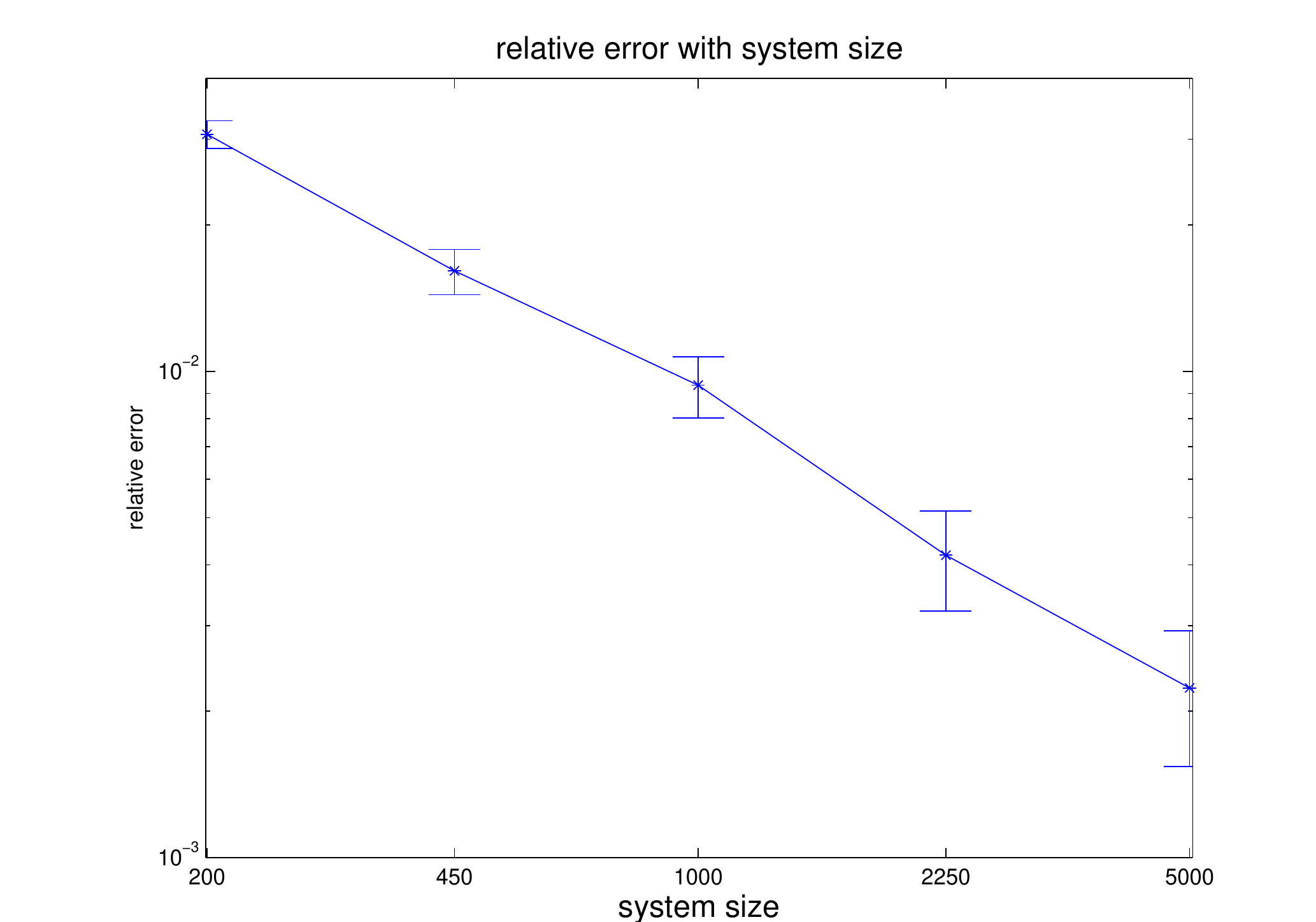}
\caption{Log-log plot of the relative deviation of finite-size simulations from cavity predictions
for 
different network sizes.}
\label{fig-N}
\end{figure}

To check the convergence of the simulation results towards cavity results valid in the thermodynamic
limit 
$N\ra\infty$, we plot the relative error of the defaulted fraction at $T=12$ compared to the cavity
simulation 
$\epsilon = \tfrac{\abs{\rho_{12}^N - \rho_{12}^\infty}}{\rho_{12}^\infty}$, for different network
sizes. 
Numerical simulations are done on networks of size $N=200$ to $N=5000$. We average over 5000 graphs
with 1000 runs on 
each graph (5000 runs for network sizes smaller than $10^3$). The results are plotted in figure
\ref{fig-N}. 
Since the variances of the cavity results are orders of magnitude smaller than those of the
simulations,
the error bars represent the RMSE of the simulations relative to the average cavity result, scaled 
by the square root of the number of simulation samples.

As expected, the finite-size simulations converge toward the message-passing solution with
increasing $N$, and the quality of the approximation is in the 1\% range even for $N$ as small as
$N\simeq 10^3$. The convergence rate is compatible with an $N^{-1/2}$ law.

\pagebreak

\section{Conclusion}

In this paper, we have studied the model of credit contagion introduced in \cite{HK06}, and have
derived an analytic solution for the damage-spreading dynamics on generic locally tree-like sparse
graphs in the thermodynamic limit, using both an adaptation of the method detailed in
\cite{Lokhov14} and a generating functional analysis. While both approaches give rise to the
\emph{same} dynamical equations, the generating functional method opens up the perspective of
developing systematic approximation schemes which could, for instance, provide simplified dynamic
equations for the large mean connectivity limit, or allow a study of finite size corrections. 
 
We have found the analytic predictions to be in good agreement with simulation results, both
for average defaulted fractions and for distributions of end-of-year defaulted fractions induced
by varying macro-economic conditions.  This comparison of simulation results for moderate system 
sizes with results that would pertain to the thermodynamic limit, therefore, shows that the present 
theoretical approach does provide a viable description of contagion dynamics in large heterogeneous 
systems, even though finite instances will contain a fair amount of short loops.

Our preliminary results for a highly schematic scenario support the idea that spillover effects
caused by asset fire-sales constitute a relevant driver of systemic instability, which appears to be
at least as important as direct contagion. However, it would be interesting to investigate the
contribution of large (scaling with $N$) numbers of assets in this model.

Our investigation of default distributions under varying macro-economic conditions shows strong risk
enhancement in the absence of countercyclical measures. 
Our analysis could be expanded to investigate the effects of such measure, for example by
considering $\xi_0$-dependent scalings of capital buffer sizes, exposure sizes (as given by
interaction weights) or of initial wealth, or by combinations of these.

The method exposed here can easily be extended to more general models. In particular, it is
straightforward to add recovery by including in our model a situation where upon default of node $i$
at time t, the loss incurred by its neighbors undergoes a damping of the type $w_{ji}(t+\tau) =
r_i(\tau)w_{ji}$, where $r_i$ is chosen among a set of damping profiles (e.g. exponential damping,
or a step function). In this case, we can make an easy parallel to the SIR model. 

We believe our approach to be particularly suited to situations where detailed information is hard to come by, or where the network is so large
 as to make direct simulation unfeasible. Indeed, the message-passing scheme presented here does not depend on the network size.


On the analytical side, open questions remain. One question of interest is the computation of large-deviation 
functions, for which an annealed computation is possible using the current method but where quenched computations run 
into difficulties (e.g. we cannot, in the replica computation, factor the bond-disorder average).

\section{Acknowledgements}

This work has been supported by the People Programme (Marie Curie Actions) of the European Unions Seventh Frame-
work Programme FP7/2007-2013/ under REA grant agreement n. 290038 (CF).

It is a pleasure to thank Luca Dall'Asta for enlightening discussions.

\appendix

\section{Appendix}
\subsection{Generating Functional Analysis}

We develop here the generating function analysis of the model described above. 
The analysis is very general and can be applied to almost any network where nodes are coupled via local fields. 
The method is well understood and has been used in a variety of other studies \cite{MC09,CR09,HWCNSC04}, although the 
version presented here differs in that instead of introducing a path integral, we can limit ourselves to a finite (low) 
number of integration parameters. 

The method proceeds as follows: considering a particular degree sequence and wealth configuration, we introduce a 
generating function for the evaluation of averages and correlation functions of observables related to contagion 
dynamics as described by eq.~\eqref{eq:1} over the graph ensemble conditioned on the degree sequence.
Taking advantage of the sparseness of the network, we express this generating function in terms of an integral with an 
effective action, and we compute the integral to order $N$ using a saddle-point approximation.

The initial part is simple: as is standard in field theory, we wish to compute a generating function for the 
correlations of observables $n_{i,t}$
\begin{equation}
G[\gv{\psi}|\v{k}.\gv{\theta}] = 
\overline{\avg{
\exp\cup{
\suml{i,t}{}\psi_{i,t}n_{i,t} 
}
}
}
\end{equation}
for a given auxiliary field $\gv{\psi}$, where the $\avg{\cdots}$ denotes an average over the dynamics (the default 
times) for a given graph and wealth realization $(\v{k},\gv{\theta})$, while $\overline{(\cdots)}$ denotes an average 
over the graphs compatible with this realization. 

Once such a function has been computed, the average value $\overline{\avg{n_{i,t}}}$ can be obtained using
\begin{equation}
\overline{\avg{n_{i,t}}} = \left.\partial_{\psi_{i,t}}G[\gv{\psi}|\v{k},\gv{\theta}]\right|_{\gv{\psi}=0} \ ,
\notag
\end{equation}
while correlation functions are given by higher derivatives, such as
\begin{equation}
\overline{\avg{n_{i,t}n_{j,t'}}} = 
\left.\partial_{\psi_{i,t}\psi_{j,t'}}G[\gv{\psi}|\v{k},\gv{\theta}]\right|_{\gv{\psi}=0} \ .
\notag
\end{equation}
In our situation, we are primarily interested in the global average
\begin{equation}
\rho(t) = \f{1}{N}\suml{i}{}\overline{\avg{n_{i,t}}} \ .
\notag
\end{equation}

Since we use the generating functional method only as a vehicle to obtain a macroscopic description
of the dynamics, we can in fact drop the source fields  $\psi_{i,t}$ entirely in what follows. 

Carrying out the average over the graphs, i.e. over their adjacency matrix $\v{A}$ and couplings 
$\v{w}=\cup{w_{ij},w_{ji}}_{(ij)}$, the generating function is expressed as
\begin{equation}
G[\v{k},\gv{\theta}] = 
\suml{\v{A}}{}\int\d{\v{w}}P(\v{A},\v{w}|\v{k})\,
\suml{\cup{t_i}}{}\prodl{i}{}
P\rp{t_i|\theta_i,\v{h}_i} \ .
\notag
\end{equation}
We then use Dirac delta functions and their Fourier representations to `extract' the dependence of
the 
$P\rp{t_i|\theta_i,\v{h}_i}$ on the couplings $\v{w}$ via the local-fields $\{\v{h}_i\}$ to obtain
\begin{align}
\label{eq:graph_average}
&G[\v{k},\gv{\theta}]= 
\suml{\v{A}}{}\int\d{\v{w}}P(\v{A},\v{w}|\v{k})\,
\int 
\prodl{i}{}
\suml{t_i}{}
\int \f{\d{\vh_i}\d{\vvh_i}}{(2\pi)^{T}}
P\rp{t_i|\theta_i,\vh_i}
\notag\\&\qquad \qquad \times
\exp \cup{-i 
\vvh_i\cdot
\rp{
\vh_i -\suml{j}{}a_{ij}w_{ij}\v{n}(t_j)
}
}\ ,
\end{align}
which, as we shall now see, allows the average over graphs to be performed.

\subsection{Graph probabilities}

When taking the average over graphs, even for tree-like graphs, we have a number of graph ensembles to choose 
from.
They fall into two broad categories: micro-canonical ensembles, where adjacency matrices are drawn so as to exactly 
reproduce a given degree sequence (with a prescribed degree distribution), and canonical models where links are 
randomly chosen such that degrees follow a given degree sequence only on average.

In the following derivation, we use a micro-canonical configuration model: we consider a "typical" (self-averaging) 
degree sequence $\v{k}$, and we take a uniform probability on the graphs with this given degree sequence:
\begin{equation}
\label{pg}
P(\v{A},\v{w}|\v{k}) \propto 
\prodl{(ij)}{}p_w(w_{ij},w_{ji})\delta_{a_{ij},a_{ji}}
\prodl{i}{}\delta_{k_i,\suml{j}{}a_{ij}} \ .
\end{equation}
It is easier for our purpose to rewrite it as
\begin{equation}
P(\v{A},\v{w}|\v{k}) \propto 
\prodl{(ij)}{}
p_w(w_{ij},w_{ji})
\delta_{a_{ij},a_{ji}}
\sp{
\f{\avg{k}}{N}\delta_{a_{ij},1}
 + \rp{1-\f{\avg{k}}{N}}\delta_{a_{ij},0}
 }
\prodl{i}{}\delta_{k_i,\suml{j}{}a_{ij}} \ ,
\notag
\end{equation}
where the extra factor is seen to be independent of the choice of $\v{A}$ for all adjacency
matrices 
compatible with the chosen degree sequence \cite{MC09}, allowing to absorb it in the overall normalization constant $\cal{N}$ of the distribution. We then use the Fourier decomposition of the Kronecker deltas to get
\begin{equation}
P(\v{A},\v{w}|\v{k}) = \f{1}{\cal{N}}
\int \prodl{i}{}\f{\d{\w_i}}{2\pi}e^{-i\w_ik_i}
\prodl{(ij)}{}p_w(w_{ij},w_{ji})
\sp{
\f{\avg{k}}{N}e^{i(\w_i+\w_j)}\delta_{a_{ij},1}
 + \rp{1-\f{\avg{k}}{N}}\delta_{a_{ij},0}
} \ .
\notag
\end{equation}
The average over weighted graphs in eq.~\eqref{eq:graph_average} factorizes with respect to the edges, so the 
generating function can be expressed as
\begin{align}
G 
&=
\f{1}{\cal{N}}
\prodl{i}{}
\suml{t_i}{}
\int
\f{\d{\vh_i}\d{\vvh_i}}{(2\pi)^{T}}
\f{\d{\w_i}}{2\pi}
e^{-i\w_ik_i}
P\rp{t_i|\theta_i,\vh_i}
e^{-i\vvh_i\cdot\vh_i}
%
\prodl{(ij)}{}D_{ij} \ ,
\notag
\end{align}
in which the individual edge contributions $D_{ij}$ take the form
\begin{align}
D_{ij} = &
\suml{a_{ij}=0,1}{}
\int \d{w_{ij}}\d{w_{ji}}p_w(w_{ij},w_{ji})
\sp{
\f{\avg{k}}{N}e^{i(\w_i+\w_j)}
\delta_{a_{ij},1}
+ 
\rp{1-\f{\avg{k}}{N}}\delta_{a_{ij},0}
}
\notag\\&\times
\exp\cup{
a_{ij}
\sp{
iw_{ij}\vvh_i\cdot\v{n}(t_j)
+
iw_{ji}\vvh_j\cdot\v{n}(t_i)
}
}
\notag\\=&
\int \d{w_{ij}}\d{w_{ji}}p_w(w_{ij},w_{ji})
\notag\\&
\times\cup{
1
+
\f{\avg{k}}{N}
\sp{
e^{i(\w_i+\w_j)}
\exp i\cup{
w_{ij}\vvh_i\cdot\v{n}(t_j)
+
w_{ji}\vvh_j\cdot\v{n}(t_i)
}
-1
}
} \ .
\end{align}
We can carry out the integration over the edge weights, and as it turns out the 
integral is factorizable even if $p(w_{ij},w_{ji})$ is not:
\begin{equation}
\label{eq:factorization}
\avg{e^{iw_{ij}\vvh_i\cdot\v{n}(t_j)}e^{iw_{ji}\vvh_j\cdot\v{n}(t_i)}}_{w_{ij},w_{ji}}
=
\avg{e^{iw_{ij}\vvh_i\cdot\v{n}(t_j)}}_{w_{ij}}
\avg{e^{iw_{ji}\vvh_j\cdot\v{n}(t_i)}}_{w_{ji}}
\end{equation}

It is intuitively clear why this should be the case: since a node once defaulted is not influenced by the subsequent 
defaults of its neighbors, the value of these couplings is irrelevant. Thus, if node $i$ defaults first, whether 
$w_{ji}$ follows the marginal distribution or the conditional $p(w_{ji}|w_{ij})$ is both irrelevant and impossible to 
determine, and we can assume the former.

From the formal point of view, this is due to a rather subtle point: for a given node $i$ with default time $t_i$, 
while the fields $\vh_i$ and $\vvh_i$ have $T$ components $(h_{i,0},h_{i,1},\cdots,h_{i,T-1})$, $P\rp{t_i|\theta_i, 
\vh_i}$ only depends on the first $t_i - 1$ 
 components of $\vh_i$. Therefore, for the remaining components, the integration 
over the $\cup{h_{i,s}}_{s\geq t_i}$ yield $\delta(\hat{h}_{i,s})$. In order to avoid cluttering the
expressions we do not 
carry out this integration explicitly, but note that $\hat{h}_{i,s}=0$ for $s\geq t_i$ implies
\begin{equation}
\exp i\cup{w_{ij}\vvh_i\cdot\v{n}(t_j) + w_{ji}\vvh_j\cdot\v{n}(t_i)}
=
\exp i\cup{
w_{ij}\suml{t_j\leq s < t_i}{}\hat{h}_{i,s}
+
w_{ji}\suml{t_i\leq s < t_j}{}\hat{h}_{j,s}
}\ ,
\notag
\end{equation}
and thus only one among the pair $(w_{ij},w_{ji})$ appears in the integral. This ``dynamical factorization'' 
is a crucial simplification. To simplify our expressions, we introduce
\begin{equation}
\chi(\vvh,t) = \avg{e^{iw\vvh\cdot\v{n}(t)}}_{w} \ .
\end{equation}

\subsection{Effective action}
Combining averages over all edge-related parts, we obtain
\begin{equation}
\prodl{(ij)}{}D_{ij} = 
\prodl{(ij)}{}
\rp{
1
+
\f{\avg{k}}{N}
\sp{
\chi(\vvh_i,t_j)\chi(\vvh_j,t_i)
e^{i(\w_i+\w_j)}
-1
}
} \ .
\notag
\end{equation}
Assuming the graph to be sparse, i.e. $\avg{k}\ll N$, this 
can be rewritten in exponential form 
\begin{align}
\label{eq:exponentiation}
\prodl{(ij)}{}D_{ij} = &
%
 \exp
\cup{
\f{\avg{k}}{N}
\suml{(ij)}{}
\sp{
\chi(\vvh_i,t_j)\chi(\vvh_j,t_i)
e^{i(\w_i+\w_j)}
-1
}
}
\notag\\ = & 
\exp
\cup{
\f{\avg{k}}{2N}
\suml{i,j}{}
\chi(\vvh_i,t_j)\chi(\vvh_j,t_i)
e^{i(\w_i+\w_j)}
-
\f{N\avg{k}}{2}
} \ .
\end{align}
We shall absorb the $\tfrac{N\avg{k}}{2}$ part into the normalization constant $\cal{N}$. 
Writing
\begin{equation}
P(\w,t,\vvh) = 
\f{1}{N}\suml{i}{}
\delta(\w-\w_i)
\delta_{t,t_i}
\delta(\vvh-\vvh_i) \,
\notag
\end{equation}
we can rewrite our previous expression \eqref{eq:exponentiation} as
\begin{equation}
\prodl{(ij)}{}D_{ij} = 
\exp
\cup{
N\f{\avg{k}}{2}
\suml{t,t'}{}
\int\d{\vvh}\d{\vvh'}
\int \d{\w}\d{\w'}
\chi(\vvh,t')\chi(\vvh',t)
e^{i\w}e^{i\w'}
P(\w,t,\vvh)
P(\w',t',\vvh')
} \ .
\notag
\end{equation}
We see that this form almost factorizes. We then introduce the quantity
\begin{align}
m(t|t') = &
\int \d{\vvh}\d{\w}
\chi(\vvh,t')
e^{i\w}
P(\w,t,\vvh)
\notag\\=&
\int \d{\vvh}\d{\w}
\f{1}{N}\suml{i}{}
\delta(\w-\w_i)
\delta_{t,t_i}
\delta(\vvh-\vvh_i)
\chi(\vvh,t')
e^{i\w}
\notag\\=&
\f{1}{N}\suml{i}{}
\delta_{t,t_i}
\chi(\vvh_i,t')
e^{i\w_i}\ ,
\notag
\end{align}
and 
decouple the sites by considering the $m(t|t')$ as integration variables, using 
auxiliary variables $\hat{m}(t|t')$ to enforce their definition via Fourier representations of 
$\delta$-functions:
\begin{align}
1 =
 \int \f{\d{m(t|t')}\d{\hat{m}(t|t')}}{2\pi/N}
 \exp\cup{
-i\hat{m}(t|t')
 \rp{
 Nm(t|t')
 -
\suml{i}{}
 \delta_{t,t_i}
 \chi(\vvh_i,t')
 e^{i\w_i}
 }
} \ .
\notag
\end{align}
We notice that after the introduction of this integral, sites are effectively decoupled. As a result the generating 
function reads, to leading order in $N$,
\begin{equation}
G
= 
\f{1}{\cal{N}}
\int  \prodl{t,t'}{}\f{\d{m(t|t')}\d{\hat{m}(t|t')}}{2\pi/N} \exp \cup{N\sp{G_1 + G_2 + G_3}} \ ,
\notag
\end{equation}
with
\begin{align}
 G_1 = &
 \f{\avg{k}}{2}
\suml{t,t'}{}
m(t|t')m(t'|t)\ ,
\notag \\
 G_2 = &
 -i\suml{t,t'}{}
 m(t|t')\hat{m}(t|t')\ ,
 \notag\\
 G_3 = &
 \f{1}{N}\suml{i}{} \ln Z_i\ ,
 \notag
\end{align}
where
\begin{align}
Z_i = &
\suml{t}{}
\int 
\f{\d{\vh}\d{\vvh}}{(2\pi)^T}
\f{\d{\w}}{2\pi}
e^{-i\w k_i}
P\rp{t|\theta_i,\vh}
e^{-i\vvh\cdot\vh}
\exp\cup{ie^{i\w}\suml{t'}{}\chi(\vvh,t')\hat{m}(t|t')} \ .
\end{align}
Using the self-averaging properties of the $(\v{k},\theta)$ configuration, we can replace $\tfrac{1}{N}\sum_i 
Z_i(k_i,\theta_i)$ with $\avg{Z(k,\theta)}_{k,\theta}$ while only making an error of order $N^{-1/2}$. 
The function $G = G_1+G_2+G_3$ is the effective action of the problem. 

\subsection{Saddle-point}
Now, considering the form of the integral, we are led in the $N\ra\infty$ limit to consider a saddle-point 
approximation, 
which will lead us to replace to leading order the integral by its value at the saddle-point. 
The saddle-point equations 
\begin{equation}
\pd{}{m(t|t')}\rp{G_1 + G_2 + G_3} = 0
\notag
\end{equation}
and
\begin{equation}
\pd{}{\hat{m}(t|t')}\rp{G_1 + G_2 + G_3} = 0
\notag
\end{equation}
read
\begin{equation}
\label{SP1}
\avg{k}m(t'|t) = i\hat{m}(t|t')
\end{equation}
and
\begin{align}
m(t|t') =&
\notag\\&
\sbox{0}{$
\f{
\int
\frac{\d{\vh}\d{\vvh}}{(2\pi)^T}
\int \frac{\d{\w}}{2\pi}
P\rp{t|\theta,\vh}
e^{-i\vvh\cdot\vh}
e^{-i\w (k-1)}
\chi(\vvh,t')
\exp\cup{ie^{i\w}\suml{t'}{}\chi(\vvh,t')\hat{m}(t|t')}
}{
\suml{s}{}
\frac{\d{\vh}\d{\vvh}}{(2\pi)^T}
\int \frac{\d{\w}}{2\pi}
P\rp{s|\theta,\vh}
e^{-i\vvh\cdot\vh}
e^{-i\w k}
\exp\cup{ie^{i\w}\suml{s'}{}\chi(\vvh,s')\hat{m}(s|s')}
}
$}\mathopen{\resizebox{1.2\width}{\ht0}{\Bigg\langle}}
\usebox{0}
\mathclose{\resizebox{1.2\width}{\ht0}{\Bigg\rangle}_{k,\theta}} \ .
\notag\\
\notag
\end{align}
We can carry out the integration over $\w$, and using \eqref{SP1}, this gives
\begin{align}
\label{eq:cond-p}
 m(t|t') =&
\notag 
\\&
\sbox{0}{$
\suml{k}{}
\f{kp_k}{\avg{k}}
\f{
\int \frac{\d{\vh}\d{\vvh}}{(2\pi)^T}
P\rp{t|\theta,\vh}
e^{-i\vvh\cdot\vh}
\chi(\vvh,t')
\cup{\suml{t'}{}\chi(\vvh,t')m(t'|t)}^{k-1}
}{
\suml{s}{}
\int
\frac{\d{\vh}\d{\vvh}}{(2\pi)^T}
P\rp{s|\theta,\vh}
e^{-i\vvh\cdot\vh}
\cup{\suml{s'}{}\chi(\vvh,s')m(s'|s)}^{k}
}
$}
\mathopen{\resizebox{1.2\width}{\ht0}{\Bigg\langle}}
\usebox{0}
\mathclose{\resizebox{1.2\width}{\ht0}{\Bigg\rangle}_{\theta}} \ .
\end{align}

We can interpret the $m(t|t')$ using the same method as in \cite{MC09}: the $m(t|t')$ \textit{at the saddle-point} 
appear as conditional probabilities, i.e. the probability that a node default at time $t$ given that we know one of its neighbors 
has defaulted at time $t'$. Thus we must have $\sum_{t}m(t|t') = 1$, 
remembering that we are also considering in this sum the probability that the node has not defaulted within the 
finite risk horizon, $1, \dots, T$. 

Another way of looking at things is that we assume the normalization $\sum_{t}m(t|t')=1$ and will
show that such solutions are self-consistent and coincide with the messages introduced 
in the message-passing solution.\\

We now remember two things. First, in the integral in the numerator of \eqref{eq:cond-p} all components $\hat{h}_s$ for $s\geq t$ cancel, as was noted previously and exploited in eq.~\eqref{eq:factorization}. 
Second, $\chi(\vvh,t')$ is a function of the scalar product $\vvh\cdot \v{n}(t') = \sum_{s=t'}^{T-1}
\hat{h}_s$. Therefore in the integral in \eqref{eq:cond-p} this sum is actually $\vvh\cdot \v{n}(t')
= \sum_{s=t'}^{t-1} \hat{h}_s$, and therefore $\chi(\vvh, t'\geq t) = \chi(0,t') = 1$.

Consequently $m(t|s\geq t) = m(t|t)$, e.g. for example $m(1|2) = m(1|1)$.  
This is clear from the interpretation of the $m(t|t')$ given above, since defaults of a neighbor after one's own default 
cannot influence the original node as was noted in previous sections.
We thus write, as previously in section~\ref{sec:MPA}, $m(t) \equiv m(t|t'\geq t)$.\\

Using these observations, we can simplify our equations further. First we notice that with these conventions, in 
eq.~\eqref{eq:cond-p} we have
\begin{equation}
\suml{t'}{}\chi(\vvh,t')m(t'|t) = \rp{1-\suml{t'<t}{}m(t')} + \suml{t'<t}{}m(t')\chi(\vvh,t')\ , 
\notag
\end{equation}
since, as was previously noted, all the components $\hat{h}_{s}$ vanish 
for $s\geq t$ and $\chi\rp{\vvh,t'}$ only depends on the components $\hat{h}_s$ for $s\geq t'$.

Second, we notice that the denominator
\begin{equation}
\notag
\suml{s}{}
\int
\frac{\d{\vh}\d{\vvh}}{(2\pi)^T}
P\rp{s|\theta,\vh}
e^{-i\vvh\cdot\vh}
\cup{\suml{s'}{}\chi(\vvh,s')m(s'|s)}^{k}
\end{equation}
 in (\ref{eq:cond-p}) is equal to 1.

Indeed, consider the simple situation where $T=2$: we have three possible trajectories:
\begin{itemize}
 \item a node defaults on the first time step ($t=1$), corresponding to the term
\[
\int
\frac{\d{\vh}\d{\vvh}}{(2\pi)^2}
W_0(-\theta)
e^{-i\vvh\cdot\vh}
 = W_0(-\theta)
 \]
 in the denominator
 \item a node defaults on the second time step ($t=2$), to which corresponds the term
 \[
\int
\frac{\d{\vh}\d{\vvh}}{(2\pi)^2}
\rp{1-W_0(-\theta)}W_1(h_1-\theta)
e^{-i\vvh\cdot\vh}
\cup{m(1)\chi(\vvh,1) + \rp{1-m(1)}}^{k}
\]
 \item a node does not default during the time horizon, to which corresponds the term
 \[
 \int
\frac{\d{\vh}\d{\vvh}}{(2\pi)^2}
\rp{1-W_0(-\theta)}\rp{1-W_1(h_1-\theta)}
e^{-i\vvh\cdot\vh}
\cup{m(1)\chi(\vvh,1) + \rp{1-m(1)}}^{k}
 \]
 as in eq.~\eqref{eq:survival}
\end{itemize}

Summing these terms, we find
\begin{align}
Z(k,\theta) = &
W_0(-\theta) + 
 \int
\frac{\d{\vh}\d{\vvh}}{(2\pi)^2}
\rp{1-W_0(-\theta)}W_1(h_1-\theta)
e^{-i\vvh\cdot\vh}
\cup{m(1)\chi(\vvh,1) + \rp{1-m(1)}}^{k}
\notag\\&
+
 \int
\frac{\d{\vh}\d{\vvh}}{(2\pi)^2}
\rp{1-W_0(-\theta)}\rp{1-W_1(h_1-\theta)}
e^{-i\vvh\cdot\vh}
\cup{m(1)\chi(\vvh,1) + \rp{1-m(1)}}^{k}
\notag\\=&
W_0(-\theta) + 
 \int
\frac{\d{\vh}\d{\vvh}}{(2\pi)^2}
\rp{1-W_0(-\theta)}
e^{-i\vvh\cdot\vh}
\cup{m(1)\chi(\vvh,1) + \rp{1-m(1)}}^{k}
\notag
\\
=&
W_0(-\theta) + \rp{1-W_0(-\theta)} = 1 \ ,
\notag
\end{align}
and the reasoning can be extended without major difficulty to $T>2$.\\

Finally, the saddle-point equations for the quantities $m(t|t')$ read
\begin{align}
& m(t|t') =
\notag\\
&
\avg{
\suml{k}{}
\f{kp_k}{\avg{k}}
\int \frac{\d{\vh}\d{\vvh}}{(2\pi)^T}
P\rp{t|\theta,\vh}
e^{-i\vvh\cdot\vh}
\chi(\vvh,t')
\cup{\suml{t''<t}{}m(t'')\chi(\vvh,t'') + \rp{1-\suml{t''<t}{}m(t'')}}^{k-1}
}_{\theta} \ ,
\notag
\end{align}
while
\begin{align}
\label{eqp1SP}
& 
m(t) =
\notag\\&
\avg{
\suml{k}{}
\f{kp_k}{\avg{k}}
\int 
\frac{\d{\vh}\d{\vvh}}{(2\pi)^T}
P\rp{t|\theta,\vh}
e^{-i\vvh\cdot\vh}
\cup{
\suml{t'<t}{}m(t')\chi(\vvh,t') + \rp{1-\suml{t'<t}{}m(t')}
}^{k-1}
}_{\theta} \ .
\end{align}
The (average) default probabilities are given by
\begin{align}
& p(t) =
\notag\\&
\avg{
\int 
\frac{\d{\vh}\d{\vvh}}{(2\pi)^T}
P\rp{t|\theta,\vh}
e^{-i\vvh\cdot\vh}
\cup{
\suml{t'<t}{}m(t')\chi(\vvh,t') + \rp{1-\suml{t'<t}{}m(t')}
}^{k}
}_{k,\theta} \ .
\notag
\end{align}
What is the connection between these equations and their message-passing equivalents ? Recall that
\[
\chi(\vvh,t) = \avg{e^{iw\vvh\cdot \v{n}(t)}}_{w} \ .
\]
Thus upon expanding \eqref{eqp1SP}, we obtain
\begin{align}
m(t) =&
\avg{
\suml{k}{}
\f{kp_k}{\avg{k}}
\int 
\frac{\d{\vh}\d{\vvh}}{(2\pi)^T}
P\rp{t|\theta,\vh}
e^{-i\vvh\cdot\vh}
\suml{q=0}{k-1}
\binom{k-1}{q}
\rp{1-\suml{t'<t}{}m(t')}^{k-q}
\right.
\notag\\
&\left.\quad\times
\suml{\tau_1,\cdots,\tau_q< t}{}
\prodl{i=1}{q}\chi(\vvh,t_i)m(\tau_i)
}_{\theta}
\notag\\=&
\avg{
\suml{k}{}
\f{kp_k}{\avg{k}}
\int 
\frac{\d{\vh}\d{\vvh}}{(2\pi)^T}
P\rp{t|\theta,\vh}
e^{-i\vvh\cdot\vh}
\suml{q=0}{k-1}
\binom{k-1}{q}
\rp{1-\suml{t'<t}{}m(t')}^{k-1-q}
\right.
\notag\\
&\left.\quad\times
\suml{\tau_1,\cdots,\tau_q< t}{}
\int \prodl{i=1}{q}\d{w_i}p_w(w_i)m(\tau_i)
e^{iw_i\vvh\cdot\v{n}(\tau_i)}
}_{\theta}
\notag\\
=&
\avg{
\suml{k}{}
\f{kp_k}{\avg{k}}
\int \d{\vh}
P\rp{t|\theta,\vh}
\suml{q=0}{k-1}
\binom{k-1}{q}
\rp{1-\suml{t'<t}{}m(t')}^{k-1-q}
\right.
\notag\\
&\left.\quad\times
\suml{\tau_1,\cdots,\tau_q< t}{}
\int \prodl{i=1}{q}\d{w_i}p_w(w_i)m(\tau_i)
\delta(\vh - \suml{i=1}{q}w_i\v{n}(\tau_i))
}_{\theta}
\notag\\=&
\suml{k}{}
\f{kp_k}{\avg{k}}
\suml{q=0}{k-1}
\binom{k-1}{q}
\rp{1-\suml{t'<t}{}m(t')}^{k-1-q}
\notag\\
&\quad \times
\suml{\tau_1,\cdots,\tau_q < t}{}
\int \prodl{i=1}{q}\d{w_i}p_w(w_i)m(\tau_i)
\avg{
P\rp{t|\theta,\suml{i=1}{q}w_i\v{n}(\tau_i)}
}_{\theta}  \ ,
\notag
\end{align}
which can be seen to be the same as \eqref{eqp1mp}.

\bibliographystyle{ieeetr}
\bibliography{./bib1}{}

\end{document}